\documentclass[12pt]{article}
\pdfoutput=1
\usepackage{amssymb}
\usepackage{graphicx}
\usepackage{graphics}
\usepackage{amsmath}

\def\d{\partial}
\def\l{\left(}									
\def\r{\right)}
\def\la{\langle }
\def\ra{\rangle }
\newcommand{\be}{\begin{equation}}
\newcommand{\ee}{\end{equation}}
\newcommand{\ba}{\begin{align}}
\newcommand{\ea}{\end{align}}
\newcommand{\bg}{\begin{gather}}
\newcommand{\eg}{\end{gather}}
\newcommand{\bseq}{\begin{subequations}}
\newcommand{\eseq}{\end{subequations}}

\renewcommand{\tanh}{\mathop{\rm th}\nolimits}

\begin{document}
\title{Scale-invariance as the origin of dark radiation?
%Evidence of scale invariance?
%Evidence for scale invariance in cosmological data?
}
\author{Dmitry Gorbunov$^{1,2}$, Anna Tokareva$^{1,3}$\\
%\affiliation{Institute for Nuclear Research of Russian Academy of
%  Sciences,\\60-th October Anniversary pr.\;7a, 117312 Moscow, Russia} 
%\affiliation{Moscow Institute of Physics and Technology,\\Institutsky
%  per.\;9, 141700 Dolgoprudny, Russia} 
\mbox{}$^{1}${\small\em Institute for Nuclear Research of Russian Academy of
  Sciences, 117312 Moscow,
  Russia}\\  
\mbox{}$^{2}${\small\em Moscow Institute of Physics and Technology, 
141700 Dolgoprudny, Russia}\\ 
%\author{Anna Tokareva}
%\affiliation{Institute for Nuclear Research of Russian Academy of
%  Sciences,\\60-th October Anniversary pr.\;7a, 117312 Moscow, Russia} 
%\affiliation{Faculty of Physics of Moscow State University,\\
%Leninskiye gory 1-2, MSU, 119991 Moscow, Russia} 
\mbox{}$^{3}${\small\em Faculty of Physics of Moscow State
  University, 119991 Moscow, Russia}
}
\date{}

\maketitle

\begin{abstract} 
Recent cosmological data favour $R^2$-inflation and some amount of
non-standard dark radiation in the Universe. We show that a framework
of high energy scale invariance can explain these data. The
spontaneous breaking of this symmetry provides gravity with the Planck
mass and particle physics with the electroweak scale. We found that
the corresponding massless Nambu--Goldstone bosons -- dilatons -- are
produced at reheating by the inflaton decay right at the amount needed
to explain primordial abundances of light chemical elements and
anisotropy of the cosmic microwave background. Then we extended the
discussion on the interplay with Higgs-inflation and on general class of
inflationary models where dilatons are allowed and may form the dark
radiation. As a result we put a lower limit on the reheating temperature in
a general scale invariant model of inflation.

\end{abstract}
%%%%%%%%%%%%%%%%%%%%%%%%%%%%%%%%%%%%%%%%%%%%%%%%%%%%%%%%%%%%%%%%%%%%%%%%%%%%

\section{Introduction}

Particle physics teaches us that in a renormalizable theory 
at high energy only dimensionless
couplings are relevant. Thus, the Standard Model (SM) becomes
scale-invariant at classical level in this limit. Though quantum
corrections generally violate scale invariance, one can speculate that
at high energy the model is indeed modified to be scale-invariant,
which provided the argument by Bardeen \cite{Bardeen:1995kv} can solve
the naturalness problem in the SM Higgs sector (suffered from the
quadratically divergent quantum corrections to the Higgs boson mass
squared, see e.g. \cite{naturalness}).  Then spontaneous breaking of
the scale invariance provides low energy particle physics with the
only (at the tree level) 
dimensionful parameter of the SM, that is the value of the
electroweak scale $v=246$\,GeV\,
\footnote{
The dark energy may be understood as either an effective cosmological
  constant emerging after spontaneous breaking of scale invariance or
  a special dynamics of dilaton field, see e.g.\,\cite{Bezrukov:2012hx}.}.

The same logic may be applied to gravity. Then at high energy the
classically scale invariant gravity 
action\footnote{Quadratic terms in the Riemann and Ricci tensors
generally  give rise to ghost-like and other instabilities and
are omitted hereafter. Since the physics responsible for violation
of the scale invariance is also beyond the scope of this paper, we
omit the dilaton potential in eq.\,\eqref{action-R2-dilaton} and
disregard its impact on the early time cosmology. Note that
the absence (smallness) of a scale-invariant quartic term $X^4$ in
\eqref{action-R2-dilaton} may be related to vanishing (tiny)
cosmological constant at later stages of the Universe expansion
\cite{Henz:2013oxa}.} 
contains both scalar curvature $R$ and dilaton $X$, 
\be
\label{action-R2-dilaton}
S_0=\int{\!\!d^4 x\, \sqrt{-g}\, \frac{1}{2}[\beta R^2+(\d_{\mu} X)^2 
- \xi X^2 R]}\,,
\ee  
with dimensionless real parameters $\beta\,,\,\xi>0$. Once the scale
invariance breaks, dilaton $X$ gains non-zero vacuum expectation
value and the last term in \eqref{action-R2-dilaton} yields the
Einstein--Hilbert low-energy action. Dilaton remains massless in
perturbation theory, so the scale invariance may be maintained at the
quantum level, see e.g. \cite{Shaposhnikov:2008xi}. As the  
Nambu--Goldstone boson, dilaton couples to other fields via derivative thus
avoiding bounds on a fifth force.   

Remarkably, with $R^2$-term in gravity action
\eqref{action-R2-dilaton}, the early Universe exhibits inflationary
stage of expansion suggested by Starobinsky \cite{starobinsky}. At
this stage the Universe becomes flat, homogeneous and isotropic as we
know it today. Also, quantum fluctuations of the responsible for
inflation scalar degree of freedom in \eqref{action-R2-dilaton}
(inflaton, which is also called {\it scalaron} in this particular model)  
transform to the adiabatic 
perturbations of matter with almost scale-invariant
power spectrum. These perturbations are believed to be seeds
of large scale structures in the present Universe and they are responsible for
the anisotropy of the cosmic microwave background (CMB). With $\beta$ 
normalized to the amplitude of CMB anisotropy $\delta T/T\sim 10^{-4}$
and $\xi X^2$ fixed by the Planck mass value in order to produce the usual gravity, the action
\eqref{action-R2-dilaton} has no free parameters. Therefore, the
inflationary dynamics is completely determined. Interestingly, recent analyses
of cosmological data \cite{Ade:2013zuv,Hinshaw:2012fq} 
favour this prediction over those of many other
models of inflation driven by a single scalar field. 

One may treat these results as a hint of scale invariance at high
energy. Yet the theory we consider apart from the Starobinsky model
contains also massless dilaton coupled to gravity through 
the last term in \eqref{action-R2-dilaton}.
In this Letter we show that this term is also responsible for the
scalaron decays into dilatons at post-inflationary reheating.  In the
late Universe the massless dilatons affect the Universe expansion.
Surprisingly, the relic amount of produced massless dilatons is precisely
what we need to explain the additional (to active neutrinos) dark
radiation component\footnote{Particular models with 
massless (Nambu--Goldstone) bosons were 
  considered in literature to address the dark radiation problem, 
see e.g. \cite{Nakayama:2010vs}.} 
suggested by the recent analyses of CMB anisotropy
data \cite{Said:2013hta, Ade:2013zuv, Hinshaw:2012fq, Sievers:2013ica,
  Story:2012wx}, and favoured by the
observation of primordial abundance of light chemical elements
\cite{Izotov:2010ca}.  We consider this finding as possibly one more hint of
scale invariance at high energy.

To complete the study we then discuss 
%in Sec.\,\ref{Sec:scalaron-vs-Higgs} 
the SM Higgs boson sector in the model 
%with scale invariance 
following Refs. 
\cite{GarciaBellido:2011de,GarciaBellido:2012zu,Armillis:2013wya} 
and outline the regions of the model parameter space where the SM
Higgs contributes to the inflationary dynamics. Finally, 
we 
investigate the dilaton production in a general scale invariant model of
a single field inflation 
%in Sec.\,\ref{Sec:general}.   
and set a lower limit on the reheating temperature from avoiding the
dilaton overproduction. 

%%%%%%%%%%%%%%%%%%%%%%%%%%%%%%%%%%%%%%%%%%%%%%%%%%%%%%%%%%%%%%%%%%%%%%%%%%%%

\section{Dilaton-scalaron inflation}
\label{Sec:Dilaton-Scalaron}

\begin{figure}[!htb]
\hskip 0.1\textwidth 
\includegraphics[width=0.8\textwidth]{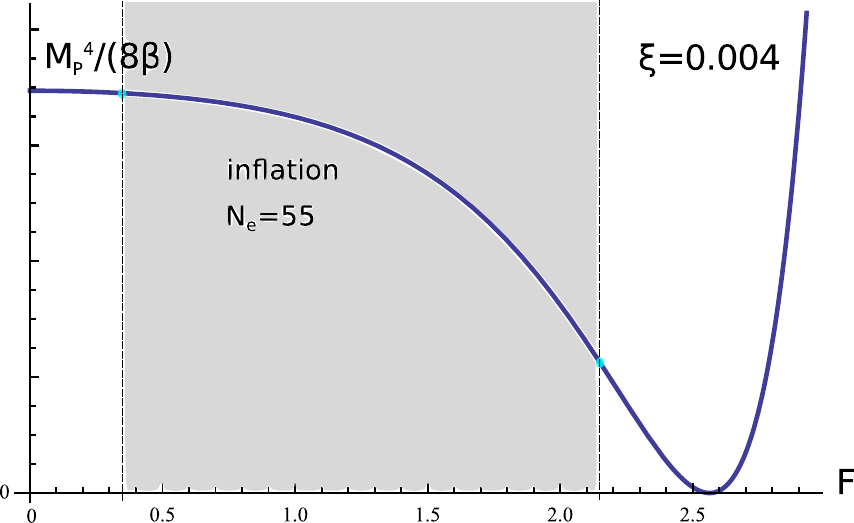} 
\caption{
Inflationary potential: field $F\equiv \l f_0 -f\r/\l M_P\sqrt{6}\r$ 
slowly moves from close to hilltop
$f\simeq f_0$ towards the minimum at $f=0$. 
The potential is symmetric under reflection $F\rightarrow -F$.}
\label{Fig:potential} 

\end{figure}

We start from considering the scale invariant extension of the
Starobinsky model with action \eqref{action-R2-dilaton}. 
  Following  \cite{Hindawi:1995cu} we introduce new scalar fields
  $\Lambda$ and ${\cal R}$ and find the equivalent form of action
  \eqref{action-R2-dilaton}:
\be
\label{Dilaton-R2-action}
S=\int{d^4 x \sqrt{-g}\, \left[\frac{1}{2}(\beta {\cal R}^2+(\d_{\mu} X)^2 - \xi X^2 {\cal R})-\Lambda {\cal R} +\Lambda R\right]}.			
\ee
Integrating out auxiliary field ${\cal R}$ (solving the
  corresponding equation of motions for ${\cal R}$) we obtain
\be
S=\int{d^4 x \sqrt{-g}\,\left[\Lambda R + \frac{1}{2}(\d_{\mu} X)^2 - \frac{1}{2\beta}(\Lambda +\frac{1}{2}\xi X^2)^2\right]}.
\ee
Going to the Einstein frame through the conformal transformation
$g_{\mu\nu}\rightarrow \tilde{g}_{\mu\nu} = \Omega^{2}g_{\mu\nu}$ with
$\Omega^2=-2\Lambda/M_P^2$, and
omitting tildes thereafter (all quantities below are evaluated with
metric $\tilde{g}_{\mu\nu}$) we arrive at 
\be 
\label{Dilaton-R-action}
S=\int{d^4 x \sqrt{-g}\, \left[-\frac{M_P^2}{2}\,R + \frac{6M_P^2}{2\omega^2}[(\d_{\mu}\omega)^2+(\d_{\mu} X)^2] - \frac{M_P^4}{8\beta}\left(1-\frac{6 \xi X^2}{\omega^2}\right)^2\right]}\;,
\ee
here $\omega=\sqrt{6} M_P \Omega$, and the reduced Planck mass $M_P$ is
defined through the Newtonian gravitational constant $G_N$ as
$1/M_P^2=8\pi G_N$. After changing the variables $\omega=r
\sin{\theta}\,, ~ X=r \cos{\theta}$ the kinetic term $K$ and potential
term $V$ become
\be 
K=\frac{6M_P^2}{2 \sin^2{\theta}}\l(\d_{\mu} \log{r})^2+(\d_{\mu}
\theta)^2\r\,,~~V=\frac{M_P^4}{8\beta}\l1-6 \xi \cot^2{\theta}\r^2\,,
\ee
or, casting them in terms of new variables
\begin{equation}
\label{new-variables}
\rho=\sqrt{6}M_P\log\frac{r}{M_P}\,,\hskip 0.5cm f-f_0
=\sqrt{6}M_P\log\tan\frac{\theta}{2}\,,
\end{equation}
we find
\be 
\label{R^2-dilaton-EF}
K=\frac{1}{2}(\d_{\mu} \rho)^2\, \cosh^2{\left(\frac{f_0-f}{\sqrt{6}M_P}\right)} +\frac{1}{2}(\d_{\mu} f)^2,~~V=\frac{M_P^4}{8\beta}\left(1-6 \xi \sinh^2{\left(\frac{f_0-f}{\sqrt{6}M_P}\right)}\right)^2.
\ee
Both kinetic and potential parts \eqref{R^2-dilaton-EF} 
are invariant under reflection $f\to 2f_0-f$. 
Choosing one of two minima of $V$ to be at $f=0$ implies that 
integration constant $f_0$ obeys 
\be 
\sinh^2\left({\frac{f_0}{\sqrt{6}M_P}}\right)=\frac{1}{6\xi}\;.
\ee
The inflation may occur at values $0<f<f_0$ (or in a mirror interval
$f_0<f<2f_0$, that we ignore in what follows), see
Fig.\,\ref{Fig:potential}. 
The potential is similar
to one considered in \cite{GarciaBellido:2011de}, so for the tilt of
scalar perturbations seeded by inflaton fluctuations one has 
\be 
n_s\simeq 1-8\xi\coth(4\xi N_e)\,,
\ee
where $N_e$ is the number of e-foldings remained till the end of
inflation from the moment when perturbations of the CMB-experiments
pivot scale $k/a_0=0.002$\,Mpc$^{-1}$ exit horizon. 
To have $N_e\approx55$ e-folds \cite{Bezrukov:2011gp} 
(since the reheating temperature is about $3.1\times 10^9$ GeV
\cite{Gorbunov:2010bn} provided scalaron decays to the Higgs bosons) 
and fit into the favoured by cosmological analyses interval 
$n_s=0.9603\pm0.0073$ \cite{Ade:2013zuv}, we need 
\begin{equation}
\xi< 0.004\;, 
\label{upper-limit-xi}
\end{equation}
hence 
$f_0>6.28\times M_P$. In order to obtain the right value of scalar
perturbation amplitude $\Delta\approx 5\times10^{-5}$ we should choose
the parameter $\beta$ (weakly depending on $\xi$) to be in the range 
$(2-0.8)\times10^9$.

%%%%%%%%%%%%%%%%%%%%%%%%%%%%%%%%%%%%%%%%%%%%%%%%%%%%%%%%%%%%%%%%%%%%%%%%%%%%%%

\section{Reheating and dilaton production}
\label{Sec:reheating}
After inflation the energy is confined in homogeneous oscillating
around minimum of the
scalaron potential. Scalaron coupling to other
fields provides oscillation decay. It reheats the Universe when the Hubble parameter becomes comparable to the inflation decay rate. 
It is well-known that scalaron couples to any {\em conformally 
non-invariant} part of the lagrangian, see discussion in   
\cite{Gorbunov:2010bn,Gorbunov:2012ij}. Within the SM the most
relevant is coupling to the Higgs field. 
Scalaron decay rate to the Higgs bosons is the same as in case of the usual
Starobinsky model, and for a more general variant with the Higgs
non-minimally coupled to gravity through the lagrangian term $-\xi'
R\, H^{\dagger}H$ one obtains \cite{Gorbunov:2012ij,Gorbunov:2012ns} 
\be
\label{Gamma_H}
\Gamma_H=\left(\frac{1}{6\beta}\right)^{3/2}\,\frac{4M_P}{192\pi}(1+6\xi')^2\;. 
\ee
Generally, scalaron decays preferably into model scalars, as their
kinetic terms are non-conformal.  				
The kinetic term in (\ref{R^2-dilaton-EF}) yields (after canonical
normalization $\rho\sqrt{1+\xi/6}\to\rho$) for
the scalaron decay rate to dilatons $\rho$ 
\be 
\label{Gamma_D}
\Gamma_{\rho}=\left(\frac{1}{6\beta}\right)^{3/2}\,\frac{M_P}{192\pi}.
\ee

We see from Eqs.\,\eqref{Gamma_H}, \eqref{Gamma_D} that 
$\Gamma_{\rho}/\Gamma_H=1/(4(1+6\xi')^2)$ giving the same ratio 
$\rho_{\rho}/\rho_{H}=1/(4(1+6\xi')^2)$ at reheating. Produced at
reheating dilatons never equilibrate in the Universe and other
mechanisms of their production (e.g. nonperturbative as discussed 
in \cite{GarciaBellido:2012zu} or in scattering of SM particles) 
are inefficient. Dilatons
contribute to the energy density and pressure of primordial plasma and
hence change the Universe expansion rate. In particular,   
the existence of the dilaton rises 
the effective number of additional to the SM 
relativistic degrees of freedom at Big Bang Nucleosynthesis \cite{GarciaBellido:2012zu}: 
\be	
\label{dark-radiation}
\Delta N_{eff}\simeq 2.85\frac{\rho_{\rho}}{\rho_{H}}=\frac{0.71}{(1+6\xi')^2}\;.
\ee

The last 9th WMAP release (more exactly, combined WMAP+eCMB+BAO+H$_0$
data) gives 
\be
\label{data}
N_{eff}=3.84\pm0.40\;. 
\ee
when helium abundance is fixed \cite{Hinshaw:2012fq}. The first result by
Planck Collaboration \cite{Ade:2013zuv} 
gives $N_{eff}=3.36\pm0.34$ in agreement with the
SM prediction $N_{eff}=3.046$. However, when independent data on 
direct measurements of the present Hubble parameter are included into
fit (which may cure the anomaly at small multipoles $l\sim 15-30$) as
was done in the WMAP result \eqref{data}, the 
estimate becomes \cite{Ade:2013zuv} (see also \cite{Said:2013hta})
\begin{equation}
\label{Planck-Neff}
N_{eff}=3.62\pm0.25\;.
\end{equation}   
Hence $\Delta N_{eff}\simeq1$ is still allowed, which is generally 
consistent with $\xi'\lesssim 1$ (when Higgs field dynamics does not
change inflation, see details in Sec.\,\ref{Sec:scalaron-vs-Higgs}).   
Moreover, one finds that for minimally coupled Higgs, $\xi'=0$, the
predicted amount of dark radiation \eqref{dark-radiation} is exactly
what we need to explain observations \eqref{data}, \eqref{Planck-Neff}.

%Planck gives $N_{eff}=3.3\pm0.27$ from the CMB \cite{Ade:2013zuv}(?).

In the conformal case $\xi'=-1/6$ or close to it, 
the Universe reheats by the anomalous inflaton decay to gauge fields \cite{Gorbunov:2012ns}. The decay rate due to the conformal anomaly is
\be 
\Gamma_{gauge}=\frac{\Sigma b_i^2\alpha_i^2 N_i}{4\pi^2} \left(\frac{1}{6\beta}\right)^{3/2}\,\frac{M_P}{192\pi}.
\ee
Here $b_i,\alpha_i, N_i$ are coefficient in $\beta$-function, gauge
coupling constant and the number of colors correspondingly for the SM
gauge fields. Numerically $\Gamma_{gauge}\sim\Gamma_{\rho}/130$ which
means that actually all inflatons decay to dilatons. 
So the case of conformal or close to conformal Higgs is forbidden. 

%%%%%%%%%%%%%%%%%%%%%%%%%%%%%%%%%%%%%%%%%%%%%%%%%%%%%%%%%%%%%%%%%%%%%%%

\section{Scalaron inflation or Higgs inflation?}
\label{Sec:scalaron-vs-Higgs}
To justify our study of the non-minimally coupled to gravity
Higgs in the context of $R^2$--dilaton inflation we need to understand 
when the nonminimal coupling $\xi'$ starts to change the inflationary dynamics. 
Consider the scale invariant action for the gravity, dilaton $X$ 
and Higgs field $h$ in the unitary gauge, 
\be
\label{Dilaton-R2-Higgs}
S_0=\int{d^4 x \sqrt{-g}\, \left[\frac{1}{2}\l\beta R^2+(\d_{\mu} X)^2 - \xi X^2 R - \xi' h^2 R+(\d_{\mu} h)^2\r - \frac{\lambda}{4}(h^2-\alpha^2 X^2)^2\right]}.
\ee	
The dilaton vacuum expectation value (vev) $\la X\ra$ defines the
reduced Planck mass (cf. Eqs.\eqref{Dilaton-R2-action} and
\eqref{Dilaton-R-action}) as $M_P=\sqrt{\xi}\la X\ra$, and the last
term in \eqref{Dilaton-R2-Higgs} defines the SM Higgs field vev as
$v=\alpha \la X\ra$. Hence the upper limit on $\xi$
\eqref{upper-limit-xi} implies $\alpha<10^{-17}$. For this
study we suppose
that at inflationary scale $\lambda>0$, which is consistent with
recent analyses \cite{Bezrukov:2012sa} 
when uncertainties are accounted for.   

Applying the same technique as in Sec.\,\ref{Sec:Dilaton-Scalaron} 
to action \eqref{Dilaton-R2-Higgs} we obtain the Einstein frame lagrangian
\be
L= -\frac{M_P^2}{2}\,R + \frac{6M_P^2}{2\omega^2}\l(\d_{\mu}\omega)^2+(\d_{\mu} X)^2+(\d_{\mu}h)^2\r - V\;,
\ee
\be
V=\frac{9\lambda M_P^4}{\omega^4}\l h^2-\alpha^2 X^2\r^2 +
\frac{M_P^4}{8\,\beta}\l 1-6\,\xi\,\frac{X^2}{\omega^2} 
-6\,\xi'\,\frac{h^2}{\omega^2} 
\r^2\;. 
\ee
The appropriate change of variables in this case looks as:
\be
\omega=r \sin{\theta}, ~ X=r \cos{\theta}\cos{\Phi}, ~ h=r
\cos{\theta}\sin{\Phi}\;.  
\ee
So we come to the lagrangian (see Eq.\,\eqref{new-variables}) 
\be 				
L=\frac{1}{2}(\d_{\mu} \rho)^2\, 
\cosh^2{F} + 
\frac{1}{2}(\d_{\mu} \phi)^2\, 
\sinh^2{F} +\frac{1}{2}(\d_{\mu} f)^2-V\,,
\ee
\be
\label{V}
V=\frac{M_P^4}{8\beta}\left[1-6  
\l\xi\cos^2{\Phi}+\xi'\sin^2{\Phi}\r\sinh^2{F} \right]^2 + 9\lambda
M_P^4\left[\l1+\alpha^2\r \sin^2{\Phi}-\alpha^2\right]^2\sinh^4{F},
\ee
where we used the following notations: $(f_0-f)/(\sqrt{6}M_P)\equiv
F$ and $\phi\equiv\sqrt{6}M_P \Phi$.  
Since $\alpha$ is expected to be tiny, its impact on inflationary
dynamics is negligible and we set $\alpha=0$ hereafter.

\begin{figure}[!htb]
%\hskip 0.1\textwidth 
\includegraphics[width=0.5\textwidth]{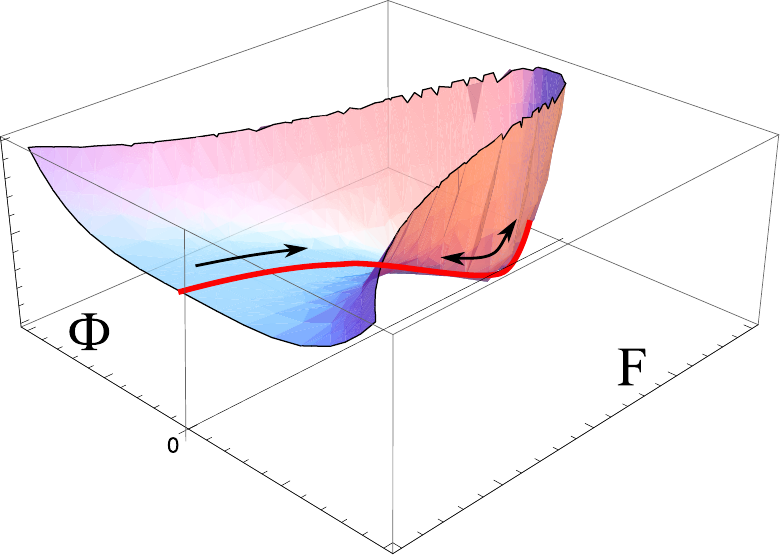} 
\includegraphics[width=0.5\textwidth]{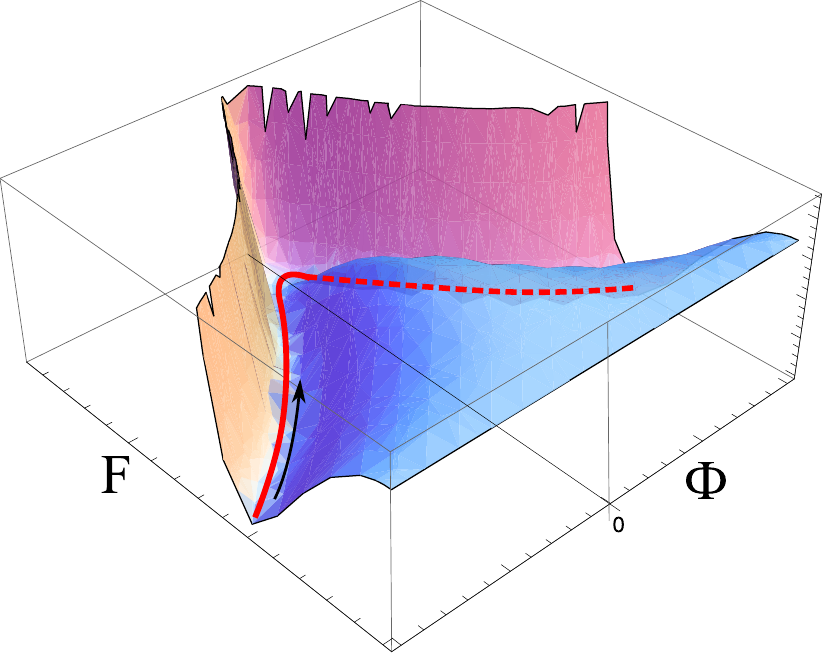}
\caption{Competition of the SM Higgs field $\phi$ and scalaron $f$ at inflationary stage. 
{\it Left plot:} $\xi'<\xi$; scalaron drives inflation and later 
reheats the Universe (like in $R^2$-dilaton inflation considered in Sec.\,\ref{Sec:Dilaton-Scalaron}). 
{\it Right plot:} $\xi'>\xi$; mostly the Higgs field drives inflation (like in Higgs-driven inflation \cite{Bezrukov:2007ep}). }
 
\label{Fig:R-H} 
\end{figure}

If $\xi'$ is small enough the situation is similar to that considered 
in Sec.\,\ref{Sec:Dilaton-Scalaron}. 
Namely, the field $f$ takes superplanckian values and 
drives slow roll inflation while the 'Higgs' $\phi$ takes small
(subplanckian) values, see the left plot in Fig.\,\ref{Fig:R-H}. 
But if $\xi'>\xi$ then $\phi=0$ is a maximum of
the potential (in $\phi$-direction) for $f>0$, see the right plot in Fig.\,\ref{Fig:R-H}, so the mentioned inflationary trajectory becomes
unstable. The stable trajectory lies in the valley described by the condition
\be 
\label{valley}
\frac{\d V}{\d \Phi}=0
\ee	
implying for the given case
\be 
\sin^2{\Phi}=\frac{\xi'-\xi}{2\beta\lambda+(\xi'-\xi)^2}\frac{1-6\xi\sinh^2{F}}{6\sinh^2{F}}\,.
\ee
The inflation along this valley exactly reproduces Higgs-dilaton
inflation \cite{GarciaBellido:2011de,Bezrukov:2007ep} for
$\xi'^2\gg\beta\lambda$, see the right plot in Fig.\,\ref{Fig:R-H}. 
In
the general case (any $\xi'^2$ and $\beta\lambda$ but with $\xi'\gg
\xi$) one observes that the kinetic term of the
field $f$ remains close to canonical when inflaton is far from its
minimum ($(1-6\xi\sinh^2{F})\sim 1$):
\be 
(\d f)^2+\sinh^2{F}(\d \phi)^2=(\d f)^2\left(1+\frac{(\xi'-\xi)\cosh^2{F}}{[1-6\xi\sinh^2{F}][12\beta\lambda\sinh^2{F}+(\xi'-\xi)(6\xi'\sinh^2{F}-1)]}  \right)
\ee
 and the effective potential along the valley \eqref{valley} is
\be 
\label{potential}
V(F)=\frac{\lambda M_P^4}{4}\frac{1}{2\beta\lambda +(\xi'-\xi)^2}\,(1-6\xi\sinh^2{F})^2.
\ee
With potential \eqref{potential} 
the amplitude of scalar perturbations is 
determined by both $\beta$ and $\xi'$, hence the latter 
may be chosen to be not as large as needed in the original
Higgs-dilaton inflation \cite{GarciaBellido:2011de}, provided the
appropriate value of $\beta$.

In all these cases the corresponding valleys attract the
inflaton trajectories: starting from general in the context of chaotic
inflation initial conditions the inflaton field approaches the
attractor and then slowly rolls along the valley. At the latter stage
the last 50-60 e-foldings happen. Then following  \cite{Kaiser:2012ak} one  
checks that neither non-gaussian nor isocurvature perturbations are
produced to be relevant in cosmology. 

To proceed with discussion of the postinflationary stage, one observes
that in all 
	these cases, $f=0$, $\phi=0$ is an absolute minimum of the
potential (\ref{V}), the inflation drives the fields towards the
origin. The expansion near this vacuum reads 
($\tilde{\phi} \equiv \phi/\sqrt{6\xi}$ is canonically normalized)   
\be
\label{V_expand}
V\approx
\l\frac{\sqrt{1+6\xi}}{\sqrt{12\,\beta}}\,M_P\, f 
+\frac{\xi'-\xi}{\sqrt{8\beta}}\,\tilde\phi^2
\r^2+
\frac{\lambda}{4}\,{\tilde\phi}^4 \,.
\ee  
At low energy field $f$ is superheavy so it decouples from the low
energy dynamic and can be integrated out leaving the SM Higgs potential.

\begin{figure}[!htb]
%\hskip 0.1\textwidth 
\centerline{\includegraphics[width=0.29\textwidth]{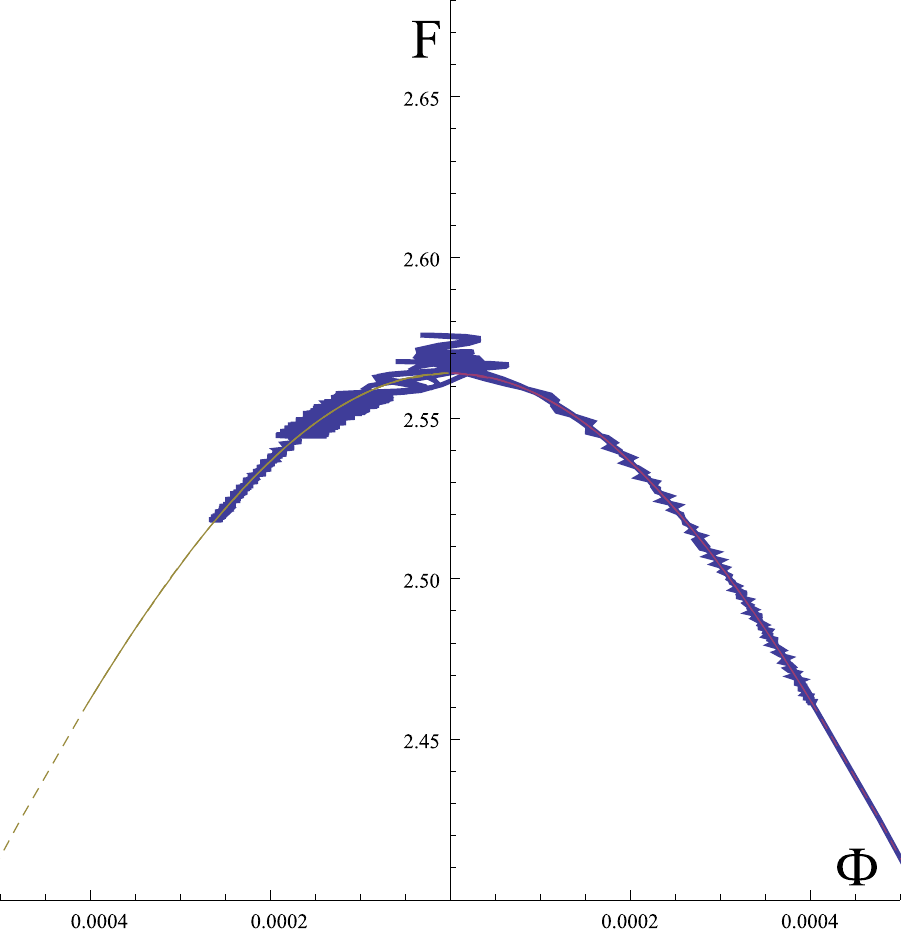}
\hskip 0.03\textwidth 
\includegraphics[width=0.31\textwidth]{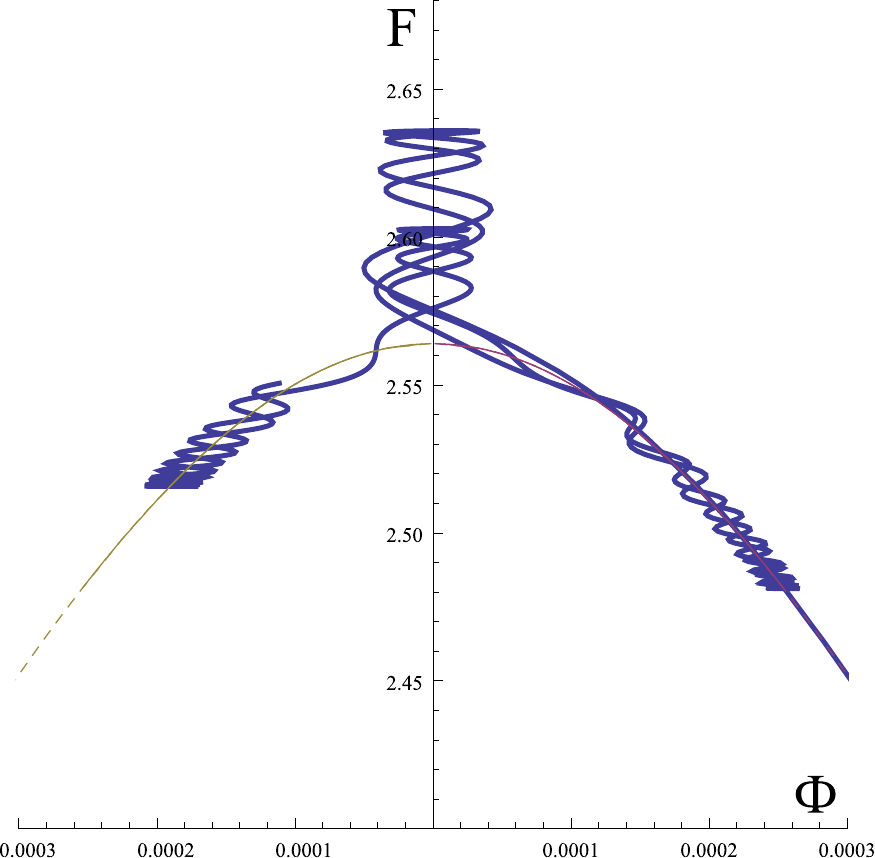}
\hskip 0.03\textwidth 
\includegraphics[width=0.31\textwidth]{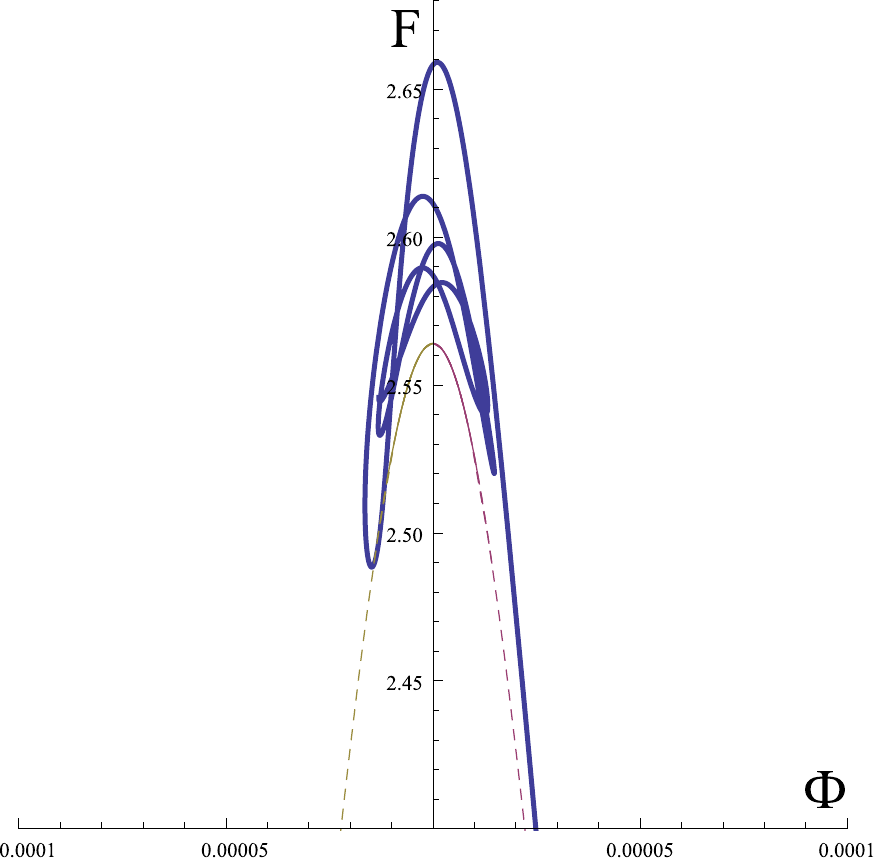}
}
\caption{Post-inflationary   
trajectories in $(\Phi,F)$ space. The dashed line corresponds to the valley \eqref{valley}. {\it Left plot:} $\xi'^2\gg
2\beta\lambda$; both inflation and
reheating as in the Higgs-inflation \cite{Bezrukov:2007ep}. 
{\it Right plot:} $\xi'^2\ll 2\beta\lambda$; Higgs-like inflation in the valley \eqref{valley} with subsequent reheating due to
scalaron decays. 
{\it Middle plot:} The intermediate case when $\xi'^2\approx 2\beta\lambda$; after inflation energy converts to both degrees of freedom.
\label{trajectories}
}
\end{figure}
\begin{figure}[!htb]
\hskip 0.1\textwidth 
\includegraphics[width=0.8\textwidth]{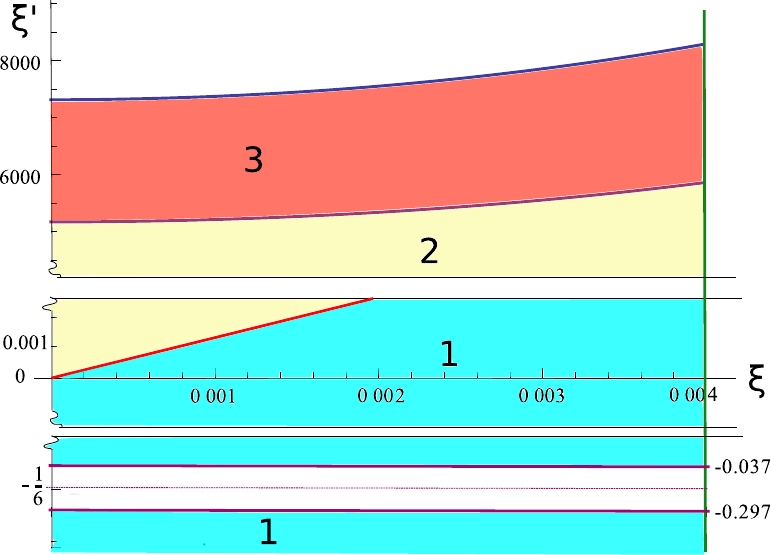} 
\caption{Shaded regions in $(\xi,\xi')$ plane are allowed from successful 
inflation and reheating. Region labeled '1' ($\xi'<\xi$) 
refers to simple scalaron inflation ended by oscillations
of the field $f$ and reheating described in
Secs.\ref{Sec:Dilaton-Scalaron} and \ref{Sec:reheating}. The region near
$\xi'=-1/6$ is forbidden because of dilaton overproduction. Domain '2'
corresponds to inflation along valley \eqref{valley} ended by oscillations
dominantly in $f$-direction and the reheating like in previous case. 
Domain '3' is for the Higgs-like inflation when subsequent 
oscillations take place
in $\phi$-direction inside the valley leading to the reheating like in
the Higgs-inflation case \cite{Bezrukov:2008ut}. 
In all these cases parameter $\beta$ is defined by
curvature perturbation amplitude $\Delta\simeq 5\times10^{-5}$, 
the e-folding number is $N_e=55$ and  
$\lambda=0.01$. }
\label{regions}
\end{figure}

The two cases mentioned above differ by direction of the inflaton
oscillations after inflation. 'Higgs'-like inflation with $\xi'\gg
\xi$ and $\xi'^2\gg 2\beta\lambda$ ends by oscillation in
$\phi$-direction which corresponds to the ordinary Higgs field, see
the trajectories on the left plot in Fig.\,\ref{trajectories}. It rapidly
decays to SM particles reheating the Universe
\cite{GarciaBellido:2012zu,Bezrukov:2008ut}.  The dilaton production
is negligible due to high reheating temperature
\cite{GarciaBellido:2012zu}. If $\xi'^2\ll2\beta\lambda$ (and
$\xi>\xi'$) the energy converts mostly to oscillations of the field
$f$, see the right plot in Fig.\,\ref{trajectories}. When $\xi'<\xi$
both inflation and oscillations take place only in $f$-direction, see
the left plot in Fig.\,\ref{Fig:R-H}, with couplings suppressed by
the Planck mass (which is similar to the Starobinsky model \cite{starobinsky})
and the reheating is delayed \cite{Gorbunov:2010bn}. The relevant for
inflation regions of model parameter space are outlined in
Fig.\,\ref{regions}.
Note that change in the reheating temperature implies (small) change in the
number of e-foldings which determines the values of cosmological
parameters (spectral indices, etc.) in an inflationary model. 

%%%%%%%%%%%%%%%%%%%%%%%%%%%%%%%%%%%%%%%%%%%%%%%%%%%%%%%%%%%%%%%%%%%%%%%%%%%%%%%

\section{Bounds on scale-invariant inflation}
\label{Sec:general}
In this section we extend our study on a general inflationary model
with scale invariance. Indeed, 
since the massless dilaton exists in all possible models with
spontaneously broken scale invariance there arises the question
whether dilaton production at reheating is high enough to give a
noticeable contribution to $\Delta N_{eff}$. Consider the
scale-invariant lagrangian for the dilaton $X$ and inflaton $\phi$
with a scale invariant
potential: 
\be 
\label{S-action}
L= -\frac{1}{2} \xi X^2 R + \frac{1}{2}(\d_{\mu} X)^2 +
\frac{1}{2}(\d_{\mu} \phi)^2 - X^4 \,V\!\!\l\frac{\phi}{X}\r.  
\ee
After conformal transformation $g_{\mu\nu}\rightarrow
\Omega^{-2}g_{\mu\nu}$ with $\Omega^2=\xi X^2/M_P^2$ and 
redefinition of fields 
\be X=\frac{r \sin{\theta}}{\sqrt{1+6\xi}}, ~
~\phi=r\cos{\theta} 
\ee 
we obtain kinetic term $K$ in the form 
\be
2 K=M_P^2 \zeta^2 \left[\frac{(\d r)^2}{r^2 \sin^2{\theta}}+\frac{(\d
    \theta)^2}{\sin^2{\theta}}\right],~~\zeta=\sqrt{\frac{1+6\xi}{\xi}}.
\ee 
Canonically normalizing the field $\theta$ and defining
$\rho=M_P\zeta\,\log{r}$ we obtain: 
\be 
\label{inflaton-dilaton}
L=\frac{1}{2}(\d
f)^2+\frac{1}{2}(\d \rho)^2\,\cosh^2{\tilde{f}} -
\frac{M_P^4}{\xi^2}\,V(\sqrt{1+6\xi}\,\sinh{\tilde{f}}).  
\ee 
Here
$\tilde{f}\equiv f/{\zeta M_P}$ and $\sin{\theta}\equiv
1/\cosh{\tilde{f}}$.   Note
that if we start from the Higgs-dilaton-like renormalizable 
potential $\lambda_0(\phi^2-\alpha^2 X^2)^2$ in the Jordan frame we
arrive at 
a potential with an exponentially flat plateau. It 
predicts close to the case of the Starobinsky model values of tilt
$n_s$ and tensor-to-scalar ratio $r$ and is  
strongly supported by the Planck data \cite{Ade:2013zuv}.

We see that the inflaton field $f$ couples to the massless dilaton
$\rho$ through its non-canonical kinetic term. So the inflaton can
decay to dilatons after inflation and produce the dark
radiation. Whether the dilaton production is negligible or not, 
depends on the function $V$ in \eqref{inflaton-dilaton}. 
Namely, if $V(y)$ has a minimum at $y=0$ then 
inflaton oscillates around the origin and the inflaton coupling to dilaton
is suppressed by $1/M_P^2$ and hence negligible. 
But if the minimum of potential is at some
nonzero $f=f_0$
the suppression factor is only $1/M_P$. Expanding around the minimum 
($f=f_0+\delta f$) we obtain in this case the interaction term 
\be
L_{int}=\frac{\tanh{\tilde{f_0}}}{\zeta M_P}\,(\d \rho)^2 \,\delta f\,,
\ee 
which corresponds to the decay width of inflaton to dilatons 
\be \Gamma_{\rho}=\frac{m^3
  \tanh^2{\tilde{f_0}}}{32\pi \zeta^2 M_P^2}\,,  
\ee 
where  $m$ is 
inflaton mass.  Requiring not to overproduce dilatons constrains
the mechanism responsible for the Universe reheating after
inflation. Namely, at reheating the dilaton production rate must be
sufficiently low as compared to the Universe expansion rate. This  
sets a lower limit on the reheating 
temperature:
\be 
T_{reh}>\frac{1.87}{\sqrt{\Delta N_{max}}}\,g_*^{-1/4}\sqrt{\Gamma_{\rho}M_P}\,,
\ee
where 
$\Delta N_{max}=N_{eff}-3.04$ is the maximal still allowed amount of
non-standard dark radiation, a rough estimate from 
\eqref{data}, \eqref{Planck-Neff} 
is $\Delta N_{max}\simeq 1$ . 

Note in passing that gravity interaction and 
scale invariance in action \eqref{S-action} supplemented with all
scale-invariant terms 
suggest two natural reheating mechanisms: decay to the SM Higgs bosons and 
anomalous decay to the SM gauge bosons (due to the conformal
anomaly). Similarly to Sec.\,\ref{Sec:reheating} we have
$\Gamma_H/\Gamma_{\rho}=4(1+6\xi')^2$. 
For the conformal or nearly conformal Higgs 
the decays into SM gauge fields dominate, so   
\be
\Gamma_{gauge}=\Sigma b_i^2\alpha_i^2 N_i \frac{m^3\,
  \tanh^2{\tilde{f_0}}}{128\pi^3 \zeta^2 M_P^2}\;, 
 \ee
adopting the same notations as in Sec.\,\ref{Sec:reheating}. 
This case is unacceptable, since  
exactly as it was obtained in Sec.\,\,\ref{Sec:reheating}, 
$\Gamma_{gauge}\sim\Gamma_{\rho}/130$, which means that mostly all inflatons
decay to dilatons grossly violating \eqref{data}. The model becomes
viable after introducing a reheating mechanism more efficient 
than the conformal anomaly.

%%%%%%%%%%%%%%%%%%%%%%%%%%%%%%%%%%%%%%%%%%%%%%%%%%%%%%%%%%%%%%%%%%%%%%%%%%%%%
\section{Conclusions}

We investigated the possibility that the probably observed
additional dark radiation has an origin associated with the 
scale invariance. Namely, the additional relativistic degree of
freedom may be massless dilaton: the Nambu--Goldstone boson of spontaneously 
broken scale invariance. Dilaton exists in all possible scale
invariant models, but its production in the early Universe and hence
its relic abundance is model-dependent. 
For example, in the Higgs-dilaton model of inflation
\cite{GarciaBellido:2011de} the dilaton gives negligible impact to the
effective number of relativistic degrees of freedom \cite{GarciaBellido:2012zu}.

We examined a natural scale invariant extension of the Starobinsky
inflationary model and found that the dilaton production in this case may be
significant and explain the observed additional dark radiation. Also we
studied the inflation and reheating taking into consideration two
fields: scalaron and Higgs in order to distinguish the parameter space
of Higgs-like inflation with negligible dilaton production and
$R^2$-like inflation giving a possibility to provide observable amount
of dark radiation. Finally, we investigated a minimal
scale-invariant extension of a single field inflation and presented
general conditions when the dilaton is produced in the amount
compatible with the recent observations.

For major part of parameter space the inflation is driven by one
  field only. The slow roll valley is an attractor, and when either
dilaton or scalaron drives inflation, the orthogonal to the inflation
trajectory direction in the field space has large mass and the valley
is deep enough, similar to what one has in case of Higgs-dilaton
inflation\,\cite{GarciaBellido:2011de}. For generic chaotic
  inflation initial conditions the 
field starts roll towards larger curvature of the potential 
 and after
 brief damped oscillations proceeds rolling inside the valley. The latest
  stage of effectively single field slow roll inflation includes the last 50-60 e-foldings of inflation
  that we can observe. Thus 
one naturally expects neither non-gaussianity nor isocurvature
perturbations at a noticeable amount 
\cite{Kaiser:2012ak,Kaiser:2013sna}. However,
in specific regions of parameter space (where both dilaton
and scalaron actively participate in observable inflationary dynamics) some
non-standard perturbations may be produced. Since both scalaron and
Higgs decays into the SM particles, the isocurvature perturbations
turn into adiabatic, which may change the amplitude of the
spectrum. Nevertheless, they may be of some interest in model
extensions, where i.e. the dark matter particles or baryon (lepton)
asymmetry are produced by scalaron or Higgs field at the reheating
stage (see e.g.
\cite{Gorbunov:2010bn,Gorbunov:2012ij,Bezrukov:2011sz}). Dilaton is
massless and its isocurvature modes would resemble those of
neutrinos. Likewise the Universe may reheat by joint work of gravity
and SM interactions, which somewhat changes the spectral indices.
Numerical calculation of these effects and of the sensitivity to the
initial (preinflationary) state we leave for future study.

\hskip 0.5cm

% {\bf Acknowledgments. } 
The authors are indebted to F.\,Bezrukov, J.\,Rubio and M.\,Shaposhnikov 
for numerous and helpful discussions. 
The work is supported in part by the grant of the
President of the Russian Federation NS-5590.2012.2 and by MSE under
contract \#8412.  The work of D.G. is supported in part by RFBR grants 
13-02-01127a and 14-02-00894a.  
The work of A.T. is supported in 
part by the grant of the President of the Russian Federation
MK-2757.2012.2.

%%%%%%%%%%%%%%%%%%%%%%%%%%%%%%%%%%%%%%%%%%%%%%%%%%%%%%%%%%%%%%%%%%%%%%%%%%%%%


\begin{thebibliography}{99}


\bibitem{Bardeen:1995kv} 
  W.~A.~Bardeen,
  ``On naturalness in the standard model,''
  FERMILAB-CONF-95-391-T.
  %%CITATION = FERMILAB-CONF-95-391-T;%%

%\bibitem{Veltman:1976rt} 
\bibitem{naturalness}
  M.~J.~G.~Veltman,
  %``Second Threshold in Weak Interactions,''
  Acta Phys.\ Polon.\ B {\bf 8}, 475 (1977); 
  %%CITATION = APPOA,B8,475;%%
%\cite{Susskind:1978ms}
%\bibitem{Susskind:1978ms} 
  L.~Susskind,
  %``Dynamics of Spontaneous Symmetry Breaking in the Weinberg-Salam Theory,''
  Phys.\ Rev.\ D {\bf 20}, 2619 (1979).
  %%CITATION = PHRVA,D20,2619;%%

%\cite{Henz:2013oxa}
\bibitem{Henz:2013oxa} 
  T.~Henz, J.~M.~Pawlowski, A.~Rodigast and C.~Wetterich,
  ``Dilaton Quantum Gravity,''
  arXiv:1304.7743 [hep-th].
  %%CITATION = ARXIV:1304.7743;%%

%\cite{Shaposhnikov:2008xi}
\bibitem{Shaposhnikov:2008xi} 
  M.~Shaposhnikov and D.~Zenhausern,
  %``Quantum scale invariance, cosmological constant and hierarchy problem,''
  Phys.\ Lett.\ B {\bf 671}, 162 (2009)
  [arXiv:0809.3406 [hep-th]].
  %%CITATION = ARXIV:0809.3406;%%
%\cite{Bezrukov:2012hx}
\bibitem{Bezrukov:2012hx} 
  F.~Bezrukov, G.~K.~Karananas, J.~Rubio and M.~Shaposhnikov,
  %``Higgs-Dilaton Cosmology: an effective field theory approach,''
  Phys.\ Rev.\ D {\bf 87}, 096001 (2013)
  [arXiv:1212.4148 [hep-ph]].
  %%CITATION = ARXIV:1212.4148;%%
  %19 citations counted in INSPIRE as of 13 Aug 2014
\bibitem{starobinsky}
A.~A.~Starobinsky,
  %``A new type of isotropic cosmological models without singularity,''
  Phys.\ Lett.\  B {\bf 91} (1980) 99;
  %%CITATION = PHLTA,B91,99;%%
A.~A.~Starobinsky, ``Nonsingular model of the Universe with the
quantum-gravitational de Sitter stage and its observational
consequences,'' in: Proc. of the Second Seminar "Quantum Theory of
Gravity" (Moscow, 13-15 Oct. 1981), INR Press, Moscow, 1982, pp.
58-72 (reprinted in: Quantum Gravity, eds. M.A. Markov, P.C. West,
Plenum Publ. Co., New York, 1984, pp. 103-128).  
 
%\cite{Ade:2013zuv}
\bibitem{Ade:2013zuv} 
  P.~A.~R.~Ade {\it et al.}  [Planck Collaboration],
  ``Planck 2013 results. XVI. Cosmological parameters,''
  arXiv:1303.5076 [astro-ph.CO].	
 

%\cite{Hinshaw:2012fq}
\bibitem{Hinshaw:2012fq} 
  G.~Hinshaw {\it et al.},
  ``Nine-Year Wilkinson Microwave Anisotropy Probe (WMAP) Observations: Cosmological Parameter Results,''
  arXiv:1212.5226 [astro-ph.CO].
  %%CITATION = ARXIV:1212.5226;%%



%\cite{Nakayama:2010vs}
\bibitem{Nakayama:2010vs} 
  K.~Nakayama, F.~Takahashi and T.~T.~Yanagida,
  %``A theory of extra radiation in the Universe,''
  Phys.\ Lett.\ B {\bf 697}, 275 (2011)
  [arXiv:1010.5693 [hep-ph]]; 
  %%CITATION = ARXIV:1010.5693;%%
S.~Weinberg, 
  %``Goldstone Bosons as Fractional Cosmic Neutrinos,''
  Phys.\ Rev.\ Lett.\  {\bf 110}, 241301 (2013)
  [arXiv:1305.1971 [astro-ph.CO]].
  %%CITATION = ARXIV:1305.1971;%%

%\cite{Said:2013hta}
\bibitem{Said:2013hta} 
  N.~Said, E.~Di Valentino and M.~Gerbino,
  ``Dark Radiation after Planck,''
  arXiv:1304.6217 [astro-ph.CO].
  %%CITATION = ARXIV:1304.6217;%%


%\cite{Sievers:2013ica}
\bibitem{Sievers:2013ica} 
  J.~L.~Sievers {\it et al.},
  ``The Atacama Cosmology Telescope: Cosmological parameters from three seasons of data,''
  arXiv:1301.0824 [astro-ph.CO].
  %%CITATION = ARXIV:1301.0824;%%

%\cite{Story:2012wx}
\bibitem{Story:2012wx} 
  K.~T.~Story {\it et al.},
  ``A Measurement of the Cosmic Microwave Background Damping Tail from the 2500-square-degree SPT-SZ survey,''
  arXiv:1210.7231 [astro-ph.CO].
  %%CITATION = ARXIV:1210.7231;%%


%\cite{Izotov:2010ca}
\bibitem{Izotov:2010ca} 
  Y.~I.~Izotov and T.~X.~Thuan,
  %``The primordial abundance of 4He: evidence for non-standard big bang nucleosynthesis,''
  Astrophys.\ J.\  {\bf 710}, L67 (2010)
  [arXiv:1001.4440 [astro-ph.CO]].
  %%CITATION = ARXIV:1001.4440;%%


  
  %\cite{GarciaBellido:2011de}
\bibitem{GarciaBellido:2011de} 
  J.~Garcia-Bellido, J.~Rubio, M.~Shaposhnikov and D.~Zenhausern,
  %``Higgs-Dilaton Cosmology: From the Early to the Late Universe,''
  Phys.\ Rev.\ D {\bf 84}, 123504 (2011)
  [arXiv:1107.2163 [hep-ph]].
  %%CITATION = ARXIV:1107.2163;%%
  %11 citations counted in INSPIRE as of 13 Mar 2013

%\cite{GarciaBellido:2012zu}
\bibitem{GarciaBellido:2012zu} 
  J.~Garcia-Bellido, J.~Rubio and M.~Shaposhnikov,
  %``Higgs-Dilaton cosmology: Are there extra relativistic species?,''
  Phys.\ Lett.\ B {\bf 718}, 507 (2012)
  [arXiv:1209.2119 [hep-ph]].
  %%CITATION = ARXIV:1209.2119;%%
%\cite{Hindawi:1995cu}

%\cite{Armillis:2013wya}
\bibitem{Armillis:2013wya} 
  R.~Armillis, A.~Monin and M.~Shaposhnikov,
  ``Spontaneously Broken Conformal Symmetry: Dealing with the Trace Anomaly,''
  arXiv:1302.5619 [hep-th].
  %%CITATION = ARXIV:1302.5619;%%

\bibitem{Hindawi:1995cu} 
  A.~Hindawi, B.~A.~Ovrut and D.~Waldram,
  %``Nontrivial vacua in higher derivative gravitation,''
  Phys.\ Rev.\ D {\bf 53}, 5597 (1996)
  [hep-th/9509147].
  %%CITATION = HEP-TH/9509147;%%
  %61 citations counted in INSPIRE as of 13 Aug 2014

  
%\cite{Bezrukov:2011gp}
\bibitem{Bezrukov:2011gp} 
  F.~L.~Bezrukov and D.~S.~Gorbunov,
  %``Distinguishing between R^2-inflation and Higgs-inflation,''
  Phys.\ Lett.\ B {\bf 713}, 365 (2012)
  [arXiv:1111.4397 [hep-ph]].
  %%CITATION = ARXIV:1111.4397;%%


%\cite{Gorbunov:2010bn}
\bibitem{Gorbunov:2010bn} 
  D.~S.~Gorbunov and A.~G.~Panin,
  %``Scalaron the mighty: producing dark matter and baryon asymmetry at reheating,''
  Phys.\ Lett.\ B {\bf 700}, 157 (2011)
  [arXiv:1009.2448 [hep-ph]].  


%\cite{Gorbunov:2012ij}
\bibitem{Gorbunov:2012ij} 
  D.~S.~Gorbunov and A.~G.~Panin,
  %``Free scalar dark matter candidates in R^2-inflation: the light, the heavy and the superheavy,''
  Phys.\ Lett.\ B {\bf 718}, 15 (2012)
  [arXiv:1201.3539 [astro-ph.CO]].
  %%CITATION = ARXIV:1201.3539;%%


\bibitem{Gorbunov:2012ns} 
  D.~Gorbunov and A.~Tokareva,
  ``$R^2$-inflation with conformal SM Higgs field,''
  arXiv:1212.4466 [astro-ph.CO].
  %%CITATION = ARXIV:1212.4466;%%


%\cite{Bezrukov:2012sa}
\bibitem{Bezrukov:2012sa} 
  F.~Bezrukov, M.~Y.~Kalmykov, B.~A.~Kniehl and M.~Shaposhnikov,
  %``Higgs Boson Mass and New Physics,''
  JHEP {\bf 1210}, 140 (2012)
  [arXiv:1205.2893 [hep-ph]]; 
  %%CITATION = ARXIV:1205.2893;%%
G.~Degrassi {\it et al.}, 
  %``Higgs mass and vacuum stability in the Standard Model at NNLO,''
  JHEP {\bf 1208}, 098 (2012)
  [arXiv:1205.6497 [hep-ph]].
  %%CITATION = ARXIV:1205.6497;%%



%\cite{Bezrukov:2007ep}
\bibitem{Bezrukov:2007ep} 
  F.~L.~Bezrukov and M.~Shaposhnikov,
  %``The Standard Model Higgs boson as the inflaton,''
  Phys.\ Lett.\ B {\bf 659}, 703 (2008)
  [arXiv:0710.3755 [hep-th]].
  %%CITATION = ARXIV:0710.3755;%%
 
 



%\cite{Bezrukov:2008ut}
\bibitem{Bezrukov:2008ut} 
  F.~Bezrukov, D.~Gorbunov and M.~Shaposhnikov,
  %``On initial conditions for the Hot Big Bang,''
  JCAP {\bf 0906}, 029 (2009)
  [arXiv:0812.3622 [hep-ph]]; 
  %%CITATION = ARXIV:0812.3622;%%
J.~Garcia-Bellido, D.~G.~Figueroa and J.~Rubio,
  %``Preheating in the Standard Model with the Higgs-Inflaton coupled to gravity,''
  Phys.\ Rev.\ D {\bf 79}, 063531 (2009)
  [arXiv:0812.4624 [hep-ph]].
  %%CITATION = ARXIV:0812.4624;%%
 
%\cite{kaiser}
 \bibitem{Kaiser:2012ak} D.~I.~Kaiser, E.~A.~Mazenc and E.~I.~Sfakianakis, %``Primordial Bispectrum from Multifield Inflation with Nonminimal Couplings,''
  Phys.\ Rev.\ D {\bf 87}, no. 6, 064004 (2013)
  [arXiv:1210.7487 [astro-ph.CO]].
  %%CITATION = ARXIV:1210.7487;%%
%13 citations counted in INSPIRE as of 22 Aug 2014



%\cite{Kaiser:2013sna}
\bibitem{Kaiser:2013sna} 
  D.~I.~Kaiser and E.~I.~Sfakianakis,
  ``Multifield Inflation after Planck: The Case for Nonminimal Couplings,''
  arXiv:1304.0363 [astro-ph.CO].
  %%CITATION = ARXIV:1304.0363;%%

%\cite{Bezrukov:2011sz}
\bibitem{Bezrukov:2011sz} 
  F.~Bezrukov, D.~Gorbunov and M.~Shaposhnikov,
  %``Late and early time phenomenology of Higgs-dependent cutoff,''
  JCAP {\bf 1110}, 001 (2011)
  [arXiv:1106.5019 [hep-ph]].
  %%CITATION = ARXIV:1106.5019;%%
  %15 citations counted in INSPIRE as of 17 Jul 2013

\end{thebibliography}
\end{document}